\documentclass[a4paper]{iopart}
\usepackage{graphicx}
\usepackage{color}
\usepackage{verbatim}
\def\gsim{\mathrel{
\rlap{\raise 0.5ex \hbox{$>$}}{\lower 0.6ex
\hbox{$\sim$}}}}
\def\lsim{\mathrel{
\rlap{\raise 0.5ex \hbox{$<$}}{\lower 0.6ex
\hbox{$\sim$}}}}

\usepackage{acronym}
\acrodef{ET}{Einstein Gravitational-Wave Telescope}
\acrodef{LIGO}{Laser Interferometer Gravitational-Wave Observatory}
\acrodef{KAGRA}{Kamioka Gravitational Radiation Antenna}
\acrodef{NS}{neutron star}
\acrodef{NS-BH}{neutron star-black hole}
\acrodef{BH}{Black hole}
\acrodef{BNS}{binary neutron star}
\acrodef{BBH}{binary black hole}
\acrodef{IMBH}{intermediate mass black hole}
\acrodef{GW}{gravitational wave}
\acrodef{GR}{general relativity}
\acrodef{SNR}{signal-to-noise ratio}
\begin{document}
\title{Scientific Objectives of Einstein Telescope}

\author{
B\,Sathyaprakash$^{26}$,
M\,Abernathy$^{1}$, 
F\,Acernese$^{2,3}$, 
P\,Ajith$^{30}$, 
B\,Allen$^{4}$, 
P\,Amaro-Seoane$^{54,32}$, 
N\,Andersson$^{5}$,
S\,Aoudia$^{54}$,  
K\,Arun$^{58}$,  
P\,Astone$^{6,7}$
% ...
}
\address{$^{1}$\,Department of Physics and Astronomy, The University of Glasgow, Glasgow, G12\,8QQ, UK}
\address{$^{2}$\,INFN, Sezione di Napoli, Italy}
\address{$^{3}$\,Universit\`{a} di Salerno, Fisciano, I-84084 Salerno, Italy}
\address{$^{4}$\,Max--Planck--Institut f\"{u}r Gravitationsphysik und Leibniz Universit\"{a}t Hannover, D-30167 Hannover, Germany}
\address{$^{5}$\,University of Southampton, Southampton SO17\,1BJ, UK}
\address{$^{6}$\,INFN, Sezione di Roma 1, I-00185 Roma, Italy}
\address{$^{7}$\,Universit\`{a} \textquoteleft La Sapienza\textquoteright, I-00185 Roma, Italy}
% ...
\address{$^{26}$\,School of Physics and Astronomy, Cardiff University, Cardiff, CF24 3AA, UK}
% ...
\address{$^{30}$\,Caltech--CaRT, Pasadena, CA 91125, USA}
% ...
\address{$^{32}$\,Institut de Ci{\`e}ncies de l'Espai, Campus UAB, Torre C-5, parells, 2na plantaES-08193 Bellaterra (Barcelona) }
% ...
\address{$^{54}$\,Max--Planck--Institut f\"{u}r Gravitationsphysik, D-14476 Potsdam, Germany}
% ...
\address{$^{58}$\,Chennai Mathematical Institute, Siruseri 603103 India}
% ...



\ead{B.Sathyaprakash@astro.cf.ac.uk}

\begin{abstract}
The advanced interferometer network will herald a new era in 
observational astronomy. There is a very strong science case to go 
beyond the advanced detector network and build detectors that 
operate in a frequency range from 1 Hz-10 kHz, with sensitivity 
a factor ten better in amplitude. Such detectors will 
be able to probe a range of topics in nuclear physics, astronomy, 
cosmology and fundamental physics, providing insights into many 
unsolved problems in these areas.
\end{abstract}

\section{Introduction}

\ac{ET} is conceived to be a third
generation detector whose conceptual design study was funded by the European 
Framework Programme FP7. The study completed in July 2011 helped produce a straw-man
design of the detector and a summary of the science (both instrumental and astrophysical) that
it promises to deliver \cite{DSD}. The accompanying article by Stefan Hild will discuss the technological
challenges and the infrastructure needed for building ET. In this article we will discuss
the rationale for going beyond advanced detectors and the huge spectrum of science and
sources that ET has the potential to uncover. 

The discussion presented here is the result of a specific study carried out 
in the context of ET. However, much of it is relevant to an extension
of either the advanced detectors or designs alternative to ET that target a 
sensitivity window from 1 Hz to 10 kHz, with the best strain sensitivity of 
$\sim {\rm few} \times 10^{-25}\,\rm Hz^{-1/2}$ in the frequency range of 20-200 Hz.  
What will ET observe in this frequency window? Why do we need detectors 
that are even more sensitive than the advanced detectors? What astrophysical problems
can be addressed with ET? These are primary questions addressed in this article.

ET, for that matter any \ac{GW} detector, is sensitive to compact objects with 
time-varying quadrupole moment.  \acp{BH} and \acp{NS}
being the most compact objects, close interactions between them, involving 
ultra-strong gravitational fields, will produce the most luminous gravitational 
radiation. ET's frequency range essentially determines the masses of compact 
objects that it could observe.  The largest angular frequency which a
BH of mass $M$ produces is roughly\footnote{We use a system
of units in which the speed of light and gravitational constant are both
equal to unity, $c=G=1.$ In this system, the mass, length and time have all
the same dimensions, taken, for convenience, to be seconds.} $\omega^2 \sim M/R^3,$ where $R$
is its size. Taking $R=2M,$ the frequency works out to be $f\sim 1.14\,\, {\rm kHz}\,
(M/10\, M_\odot)^{-1}.$ For comparison, the most dominant quasi-normal mode
frequency of a $10\,M_\odot$ Schwarzschild BH is 1.19 kHz and that of a
Kerr BH (with dimensionless spin of 0.9) is 2.15 kHz. Thus, the
frequency range of 1-$10^4$ Hz gives a mass range of $1$-$10^4\,M_\odot.$ 

It might at first appear that the low-frequency window of 10-20 Hz, where
the noise floor could be an order of magnitude or two larger than at 20-200 Hz, has no
particular advantage for enhancing the visibility of signals. This is possibly true 
in the case of sources that sweep past the best part of the detector sensitivity.
However, good low-frequency sensitivity does two things: Firstly,
opens up a window for observing \acp{IMBH} \cite{Gair:2010dx,AmaroSeoane:2009ui}
with masses in the range $10^3$-$10^4\,M_\odot.$ There is as yet no conclusive
evidence for the existence of IMBH, let alone their
binaries. However, there are strong indications that certain ultra-luminous x-ray
sources (e.g.\, HLX-1 in ESO 243-49 \cite{2009Natur.460...73F}) are host to 
IMBH.  If a population of such objects exists and
they grow by merger, then, depending on their masses,  ET will be able to 
explore their dynamics out to $z\sim6$-15 and study their mass function,
redshift distribution and evolution.  Secondly, lower frequencies
help improve measurement accuracies of source parameters. Binary systems spend
very long periods at lower frequencies, with the time to coalescence from a frequency
$f$ rising as $f^{-8/3}.$ The long duration over which the sources slowly chirp-up 
in frequency helps in measuring the parameters of the source very accurately. For 
instance, in the case of advanced LIGO, the \ac{SNR} for a \ac{BNS} 
signal integrated from 10 Hz to 20 Hz is less than 1\% of the 
SNR that accumulates above 20 Hz until merger. Yet, the measurement accuracy of the system's
masses is a factor of two better if the signal is integrated from 10 Hz instead of 20 Hz.  
This effect will be even stronger in the case of ET as its lower frequency cutoff 
could be a factor ten smaller compared to advanced detectors.

The population of sources in the frequency window from 100 Hz to 10 kHz 
is also known to be very rich and there are many challenges and opportunities 
in this frequency region, both in instrument design and astrophysical potential.
Quakes in \acp{NS} (believed to be the root cause of glitches in radio
pulsar observations), giant explosions that occur in magnetars, gravitational
collapse and supernovae, dynamics of accreting \acp{NS}, relativistic instabilities 
in young and accreting \acp{NS}, are all potential sources where observations could reveal
a wealth of information that is complementary to radio, x-ray or gamma-ray observations.

We will begin the discussion with a brief recap of what we can expect from
a network of advanced detectors over the next decade. We will then go on to
describe the topology and sensitivity of ET and why ET has additional 
advantages over equivalent L-shaped detectors. This is followed by a list of
sources that ET can observe and how that benefits in furthering our
knowledge of fundamental physics, cosmology and astrophysics. 

\begin{figure*}[t]
\centering
\includegraphics[width=24pc]{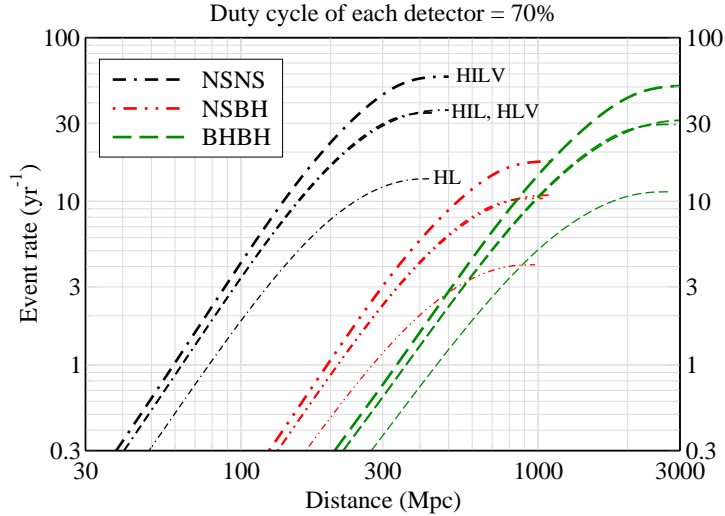}
\caption{\label{fig:network rate}The plot shows the cumulative 
number of compact binary events expected  to be detected by a 
network within a given distance, for three archetypal 
compact binaries and four different advanced detector networks. The
curves flatten (and stay constant) upon reaching the {\em horizon 
distance} of the network, the distance beyond which a network
cannot detect signals with the desired signal-to-noise ratios. 
See the text for further details.
}
\end{figure*}
\section{Advanced gravitational wave detectors}

Advanced interferometric gravitational-wave detectors (advanced \ac{LIGO}
\cite{Smith2009,advLIGO:2007} in the US, advanced Virgo \cite{0264-9381-23-19-S01} 
in Italy and \ac{KAGRA} \cite{LCGTProject} in Japan) will be built and 
become operational over the next 3 to 5 years.
As discussed by Alan Weinstein in this issue, the global network of advanced 
detectors is expected to open the gravitational window for observational astronomy.
\ac{BNS} mergers are the prime candidate sources for advanced
detectors. Extrapolation of the galactic \ac{BNS} population to extragalactic Universe
suggests that we may nominally observe one merger event every week 
\cite{Abadie:2010cf}.  Figure \ref{fig:network rate} plots the
expected number of compact binary mergers detectable within a given
volume by the advanced detector network of LIGO and Virgo, for three 
archetypal compact binaries and four different detector networks.

The various networks considered are: (1) a two-detector network
consisting of LIGO Hanford and LIGO Livingston (HL) detectors, 
(2) two three-detector networks, consisting of HL and either Virgo 
(HLV) or LIGO India\footnote{LIGO is currently considering moving one of the 
Hanford detectors to India. However, the event rates shown here
do not depend on whether the third LIGO detector is relocated to
India or remains at Hanford.} (HIL) and (3) the full four-detector 
network of three LIGO detectors and Virgo (HILV).  Neutron stars are 
assumed to be 1.4 $M_\odot$ and \ac{BH} $10\,M_\odot.$ 

A signal is said to be detectable by a detector network if it 
produces an SNR of 5 or more in at least two detectors and a network SNR 
$\ge 12.$ Note that the criteria used in computing the detection rate here
is somewhat different from that used in Ref.\,\cite{Abadie:2010cf}. 
In computing the detection rate we have taken the local 
merger rate of \ac{BNS}, \ac{NS-BH} and \ac{BBH} sources, in a volume of $10^6\,\rm Mpc^{3},$ 
to be $1\,\rm yr^{-1},$ $0.03\,\rm yr^{-1}$ and $5\times 10^{-3}\rm yr^{-1},$ 
respectively \cite{Abadie:2010cf}, and included a factor of $(1+z)^{-1}$ 
to account for the reduction in the rate due to cosmological expansion.  
The expected detection rate (per year) also takes into account the varying 
efficiency of the detector as a function of distance, the live-time of the different 
networks (assuming a duty cycle of 70\% for each detector in a network) 
and ``redshift" effect of binary masses, by which observed masses 
are larger by a factor $(1+z)$ relative to intrinsic masses. The
redshift of masses has a significant effect on the reach of the
network for \acp{BBH}, leading to a 30\% increase in the reach and a 
factor of two in the event rate compared to numbers quoted in
Ref.\,\cite{Abadie:2010cf}. The detection rate increases by a factor 
of three as we go from two to three detectors and by little less 
than a factor of two as we go from three to four detectors. 

The point at which the various curves level off is the 
{\em horizon} of the network -- the maximum distance
up to which events can be detected in that network. For
instance, the reach for \ac{BNS} sources is about $\sim 500$ Mpc for 
all networks considered and the expected event rate per 
year within this volume is 13, 33 and 60 in HL, HIL/HLV and
HILV networks, respectively. 
The uncertainty in the expected rate of \ac{BNS} mergers
is so large that the detected number of events 
could be several per day to once every two years.
At any rate, the advanced detector network will make the
first direct detection of \acp{GW} within this decade.

The immediate consequence of the first observations (of \ac{BNS} systems) is to pin
down the merger rate to within an order of magnitude. If the actual rate is
close to the mode of the distribution of predicted rates, then there will be occasional
events with large enough SNR to measure the mass and radius of \acp{NS} and 
constrain their equation of state \cite{Read:2009bns,2011GReGr..43..409A}. 
The observed population will also give us the mass function of \acp{NS} in 
binary systems, which is important for testing models of compact binary 
formation and evolution. 

Advanced detectors might also verify if compact binaries, in which at least one
of the component objects is a \ac{NS}, and the other either a \ac{NS} or a 
\ac{BH}, are progenitors of short-hard gamma ray bursts (shGRBs) 
\cite{Nakar2007}.  The observed rate of shGRBs suggests that their rate could 
be once per year within $z\sim 0.5$ \cite{Dietz:2010eh}.
The reach of the network can be significantly larger for
a search that is focussed around an event like a GRB, than it is
for a ``blind" search in the entire data set. This is largely because
the duration over which the search needs to be carried out will
be seconds, instead of years, and so it is possible to detect
events with a network SNR of about 6, instead of 12 \cite{Harry:2010fr}. 
In the case of GRBs, the reach for \ac{BNS} is about 1 Gpc and for NH-BS 
is $z\simeq 0.4.$ Thus, within 5 years of operation, advanced detector network 
should confirm whether \ac{BNS} or \ac{NS-BH} mergers are progenitors of shGRBs. 
However, it is unlikely that the advanced network will be able to
carry out a detailed study of different types of GRBs, the relationship between
their duration, spectra, and demographics.

The advanced network has a great discovery potential. For instance, they could observe, 
for the first time, populations of \ac{NS-BH} binaries and \acp{BBH}. Although
it is largely expected that radio, x-ray and gamma-ray observations could soon discover
a \ac{NS-BH} system, observation of a stellar-mass \ac{BBH} is most likely to come
from \ac{GW} observations.  
If the rate of \ac{BBH} mergers in the Universe is $\sim 5\times 10^{-9}\,\rm yr^{-1}\,Mpc^{-3},$
then we might see as many as $\sim 60$ events per year (see Figure \ref{fig:network rate}).
At the lower end of the merger rate ($10^{-10}\,\rm yr^{-1}\,Mpc^{-3}$), the observed number 
of signals could still be $\sim 1$ per year.  
Some authors predict far higher rates \cite{Belczynski:2011qp}, in which case
the advanced detector network will routinely observe many bright \ac{BH} mergers.
Whatever the rate, the advanced \ac{GW} network 
is expected detect first \acp{BBH} and constrain astrophysical models of
their formation and evolution, as also indirectly 
constraining the metallicity of gas as a function of redshift \cite{Belczynski:2010tb}.

Advanced detectors could also observe weak galactic sources such as
\acp{NS} with mountains that have an effective ellipticity of \cite{DSD}
$\epsilon \sim 10^{-9}$-$10^{-7}$ (with spin frequencies in the range 0.1-1 kHz), 
an occasional supernova and other galactic sources,
if they occur in sufficient strength and numbers. 
Advanced detectors could provide a handful of sources with moderately high \ac{SNR} 
$(\gsim 50)$ events, but they are unlikely to yield a great number of sources, nor
sources with very large \ac{SNR} -- essential requisites for precision astronomy. 
To do so would necessitate a detector that has improved low-frequency 
sensitivity and greater amplitude sensitivity, a factor of 10 in both.

\section{Beyond the advanced detector network}
The rich variety of sources (rich, both in terms of the 
types of sources as well as their spectra) that a \ac{GW} detector
promises to observe, opens the possibility of using the \ac{GW} window for
furthering our understanding of fundamental physics, cosmology and
astrophysics. Detailed study of individual sources, e.g. \ac{NS} cores, \ac{BH}
quasi-normal modes, is only possible if the SNR is in excess of $\sim 50.$
The same is true for strong field tests of gravity which would benefit 
from loud events with SNR $ \gg 50.$ 

However, loud events are not the only reason why we need to go beyond the
advanced network. Let us look at an example why this is so.  Compact binaries 
are standard sirens that could be used to precisely measure the luminosity
distance to a source without the aid of any cosmic distance ladder. Additionally,
if one can also identify their host galaxies and measure their redshifts,
then one could infer the cosmological parameters. For accurate measurement
of cosmological parameters it is not sufficient to have a few loud sources.
This is because gravitational weak lensing of cosmological sources could
bias the luminosity distance of a source, the systematic error being as 
large as 5\% at a redshift of $z=1.$ It is not possible to correct for such 
errors \cite{ShapiroEtAl09},
but it is possible to mitigate the effect of weak lensing if the
number of sources is $\sim 400.$ Since gravitational lensing can cause
the distance to be under- or over-estimated, the availability of a large
number of sources helps in statistically nullifying the bias. 

A network of detectors with ten times better amplitude sensitivity and a factor 
ten reduction in the low-frequency seismic floor compared to advanced detectors, 
can help explore stellar-mass \acp{BH} when the Universe was still assembling its first 
galaxies, discover binaries consisting of \acp{IMBH} out to
a redshift of $z\simeq 6,$ \cite{Gair:2010dx,AmaroSeoane:2009ui} detect 
every shGRB within a redshift of 
$z\simeq 4$ assuming \ac{BNS} or \ac{NS-BH} are their progenitors and observe a variety 
of weaker galactic sources such as \acp{NS} with ellipticity larger 
than ${\rm few}\times 10^{-8}$-${\rm few}\times 10^{-10}$, glitching pulsars 
in the Milky Way that deposit more than $10^{-12}\,M_\odot$ of rotational 
energy in GW \cite{2011GReGr..43..409A}, magnetars within Andromeda that 
emit $\sim 10^{-8}\,M_\odot$ in GW \cite{ChassandeMottin:2010zh}
and core-collapse supernovae that occur within 2-4 Mpc.  Such a detector 
is equivalent to going from a 1 m class optical telescope 
to a 100 m class telescope but also extending the observation to infra-red
frequencies. The massive scientific potential makes a very compelling case to 
go beyond the advanced detector network. 

While a single-site third generation
detector might achieve some of the scientific goals discussed in this review,
any problem that necessitates a knowledge of the position of the source on 
the sky and its distance from the Earth will require a network of detectors.
For example, while strong field tests of \ac{GR} could be performed with
events detected in a single-site ET, measurement of cosmological parameters
would require a network of detectors. Likewise, testing the propagation speed
of gravitational waves relative to electromagnetic waves from a supernova,
would not require a network of detectors; optical identification of the supernova
and coincident detection of gravitational waves in a single ET, could
confirm if gravitons are massive. One should also be mindful of the 
covariance between various parameters of a source before deciding if it 
is safe to draw scientific conclusions based on observations in a single-site ET.

The rest of this Section is organized as follows: We will first take a 
look at the topology of ET and its advantages over the conventional 
L-shape configuration. We will then review ET's sensitivity and its ability
to detect compact binaries and other sources. 
We will conclude with the data analysis challenges posed by ET
and efforts to tackle signal discrimination and measurement problems. 

\subsection{Topology}
The ET design study team concluded that a triangular topology 
is the optimal strategy to achieve the sensitivity goal 
of a third generation detector \cite{Freise:2008dk,Freise:2009nz}. 
\begin{figure*}[t]
\centering
\includegraphics[width=24pc]{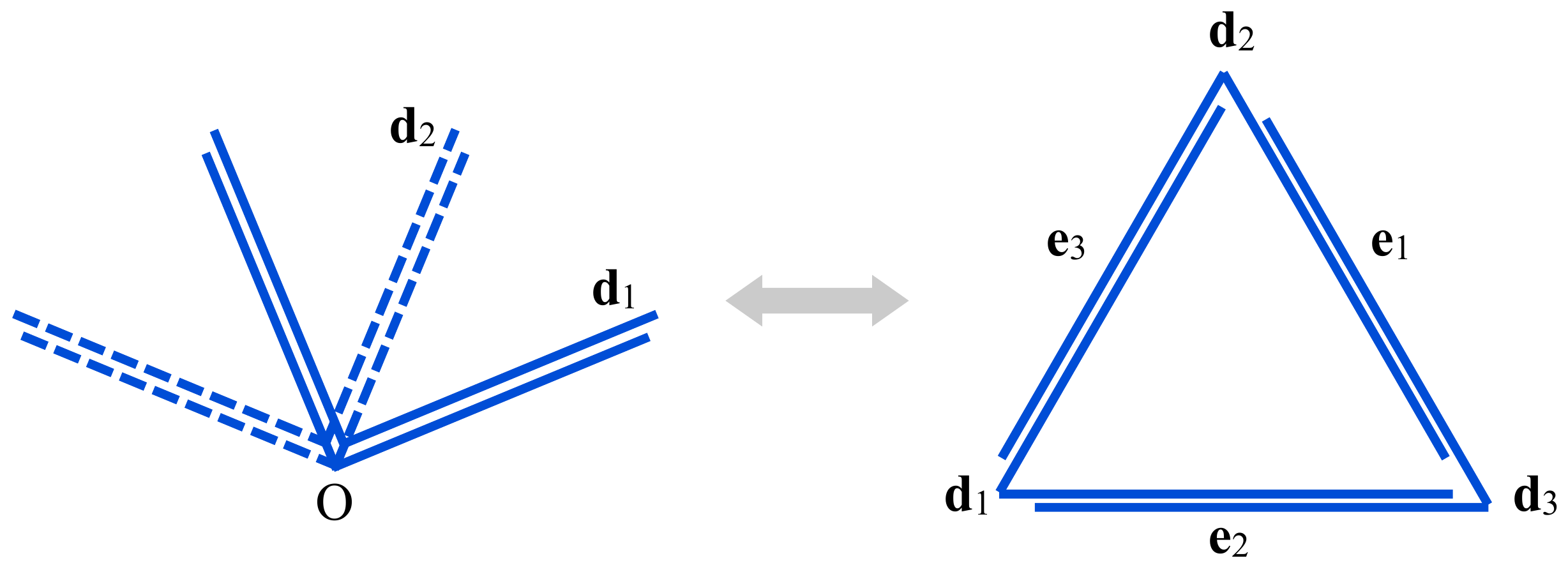}
\caption{\label{fig:topology}Triangular topology of ET, 
in which each arm of length $L$ is used twice to form three
detectors with a 60-degree opening angle, is equivalent to that of two 
L-shaped detectors of length $3L/4,$ whose arms house two detectors in
each \cite{Freise:2008dk}. (Arms are drawn to scale.)}
\end{figure*}
The arms of the triangle are each used twice to form three Michelson 
interferometers as shown in Figure \ref{fig:topology}. Each V-shaped
detector in the array has $L=10$~km arms, with an opening angle of $\alpha=60$ 
degrees and the detectors are rotated relative to each other by an angle of 
120 degrees.  One way to characterize the topology is to look at the 
antenna pattern of the network.  The pattern functions $F_+$ and 
$F_\times$ of each detector in the ET array are identical to that of an 
L-shaped detector with arm length
$L\sin^2\alpha=7.5\,\rm km.$ Figure \ref{fig:antenna
pattern} compares the quadrature sum $(F_+^2+F_\times^2)^{1/2}$
of Virgo (which is an L-shaped interferometer) to the {\em joint}
antenna pattern of the three ET interferometers located at the same site as Virgo. 
ET has essentially no blind spots on the sky \cite{DSD,Regimbau:2012ir}.  
In a coordinate system in which the three arms of ET lie in the $xy$ 
plane, the array is insensitive to the source's azimuth angle. 

The three detectors of the ET array are equivalent, in terms of the 
antenna pattern and sensitivity, to two L-shaped detectors whose arms 
are three-quarters in length and rotated relative to each other by 
an angle of 45 degrees \cite{Freise:2008dk}. ET will have
fewer end stations than the design with two L-shapes.
ET's three interferometers allow the construction of a {\em null}
data stream \cite{Wen:2005ui} that is completely devoid of \acp{GW} 
\cite{Freise:2008dk,Regimbau:2012ir}, as do a pair of L-shaped detectors
sharing the two arms. Finally, each
pair of ET interferometers can solve for the plus and cross polarizations.

\subsection{Null data stream and polarization of gravitational radiation}
For a triangular detector (for that matter for any closed topology), 
the sum of the responses contains only the sum of the background 
noise from the different interferometers and no \ac{GW}
signal \cite{Freise:2008dk}. 
Let us denote by $h^A(t),$ $A=1,2,3,$ the response functions
of the three interferometers in the ET array. By definition
$$h^A(t)=F^A_+\,h_+ + F^A_\times\,h_\times,$$
where $h_+$ and $h_\times$ are the plus and cross polarizations 
of the incident signal and $F^A_{+}$ and $F^A_{\times}$ are the plus and 
cross antenna pattern functions. The plus and cross pattern functions are 
inner products of the detector tensors $d^{ij}_A$ and the polarization 
tensors $e^+_{ij}$ and $e^\times_{ij},$ respectively:
$$
F_+^A=d_A^{ij}\, e^+_{ij},\quad F_\times^A=d_A^{ij}\, e^\times_{ij}.
$$
Since the detector tensors are given by 
$$d_1^{ij}=\frac{1}{2} \left ( e_2^i\, e_3^j - e_3^i\, e_2^j\right ),\quad
  d_2^{ij}=\frac{1}{2} \left ( e_3^i\, e_1^j - e_1^i\, e_3^j\right ),\quad
  d_3^{ij}=\frac{1}{2} \left ( e_1^i\, e_2^j - e_2^i\, e_1^j\right ),
$$
where ${\mathbf e}_1,$ ${\mathbf e}_2$ and ${\mathbf e}_3,$ are unit vectors 
along the arms of the detectors as in Figure \ref{fig:topology}.
It is easy to see that $\sum_A h^A=0,$ irrespective of the direction
in which the radiation is incident or its polarization. Thus, the
sum of the detector outputs $\sum_Ax^A(t)=\sum_A [n^A(t)+h^A(t)]=\sum_A n^A(t),$
contains only the sum of the three noise backgrounds.
This is the {\em null data stream,} which is completely
devoid of any \acp{GW}. It is like the dark-current
of an optical telescope. It can be used to veto out spurious events \cite{Wen:2005ui}, to 
estimate the noise spectral density of the detectors (which is critical for a 
signal-dominated instrument such as ET) and to detect a stochastic
\ac{GW} background that may be buried in the data streams \cite{Regimbau:2012ir}.

\begin{figure*}
\centering
\includegraphics[width=0.75\textwidth]{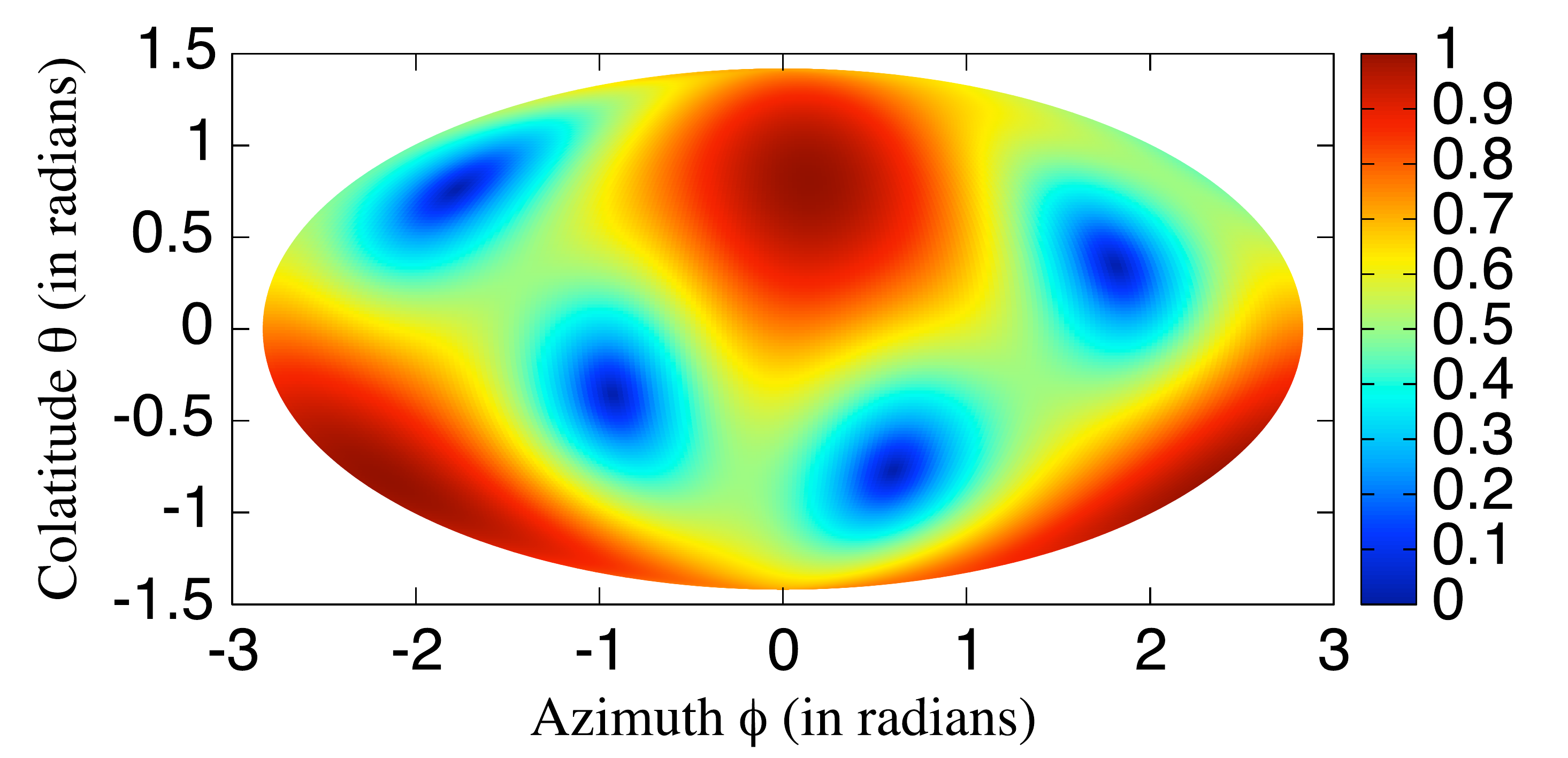}
\includegraphics[width=0.75\textwidth]{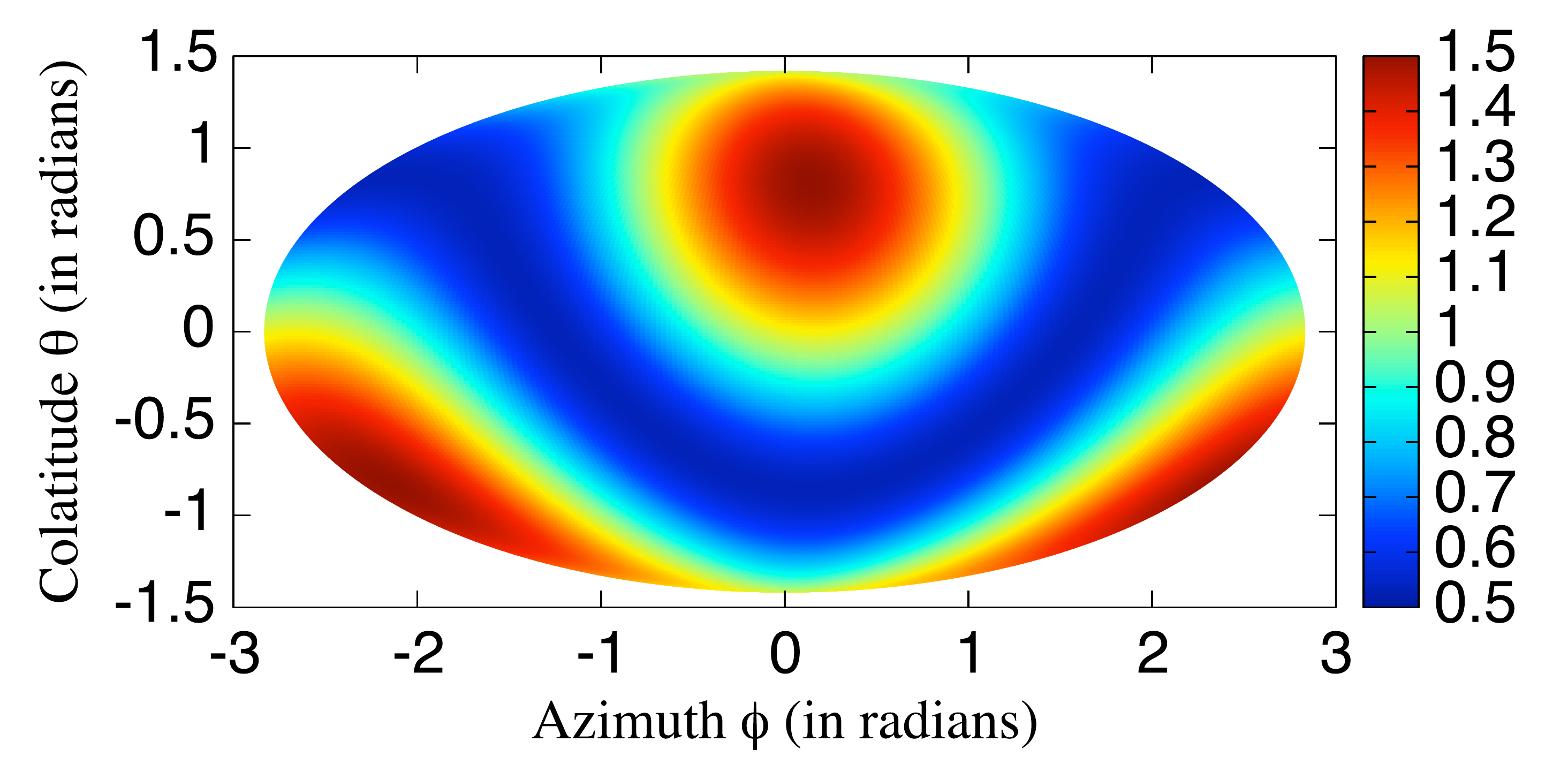}
\caption{\label{fig:antenna pattern}Antenna pattern of the Virgo interferometer
(top) compared to that of ET (bottom) located at the same site.}
\end{figure*}

ET has also the ability to resolve the wave's two polarization states.
It is straightforward to invert ET's response functions
$h^A=F_+^A\,h_+ + F_\times^A\,h_\times,$ $A=1,2,3,$ to solve for the
signal's two polarizations. We won't have direct access to the
$h^A$ but only to detector outputs $x^A=h^A+n^A$ whose linear
combinations could be used to get estimates of the two polarizations.

For a pair of misaligned L-shaped detectors (i.e., detectors that are
not rotated relative to each other by multiples of $\pi/2$ radians)
there is no linear combination of their responses that gives a null data 
stream. Nevertheless, as noted before, two interferometers sharing common
arms (as in the LIGO Hanford setup) can be used to construct a sky-independent
null stream. Even so, a triangular topology can provide
the same information and yet incur lower infrastructure costs.  If one is 
going to build two pairs of L-shaped detectors then it makes sense to 
build them at geographically widely separated sites, as that would help 
in obtaining at least partial information about source position, which would 
obviously increase the science reach over a single-site ET.

\subsection{ET's sensitivity}

\begin{figure*}
\begin{center}
\hskip-1cm
\includegraphics[width=0.69\textwidth]{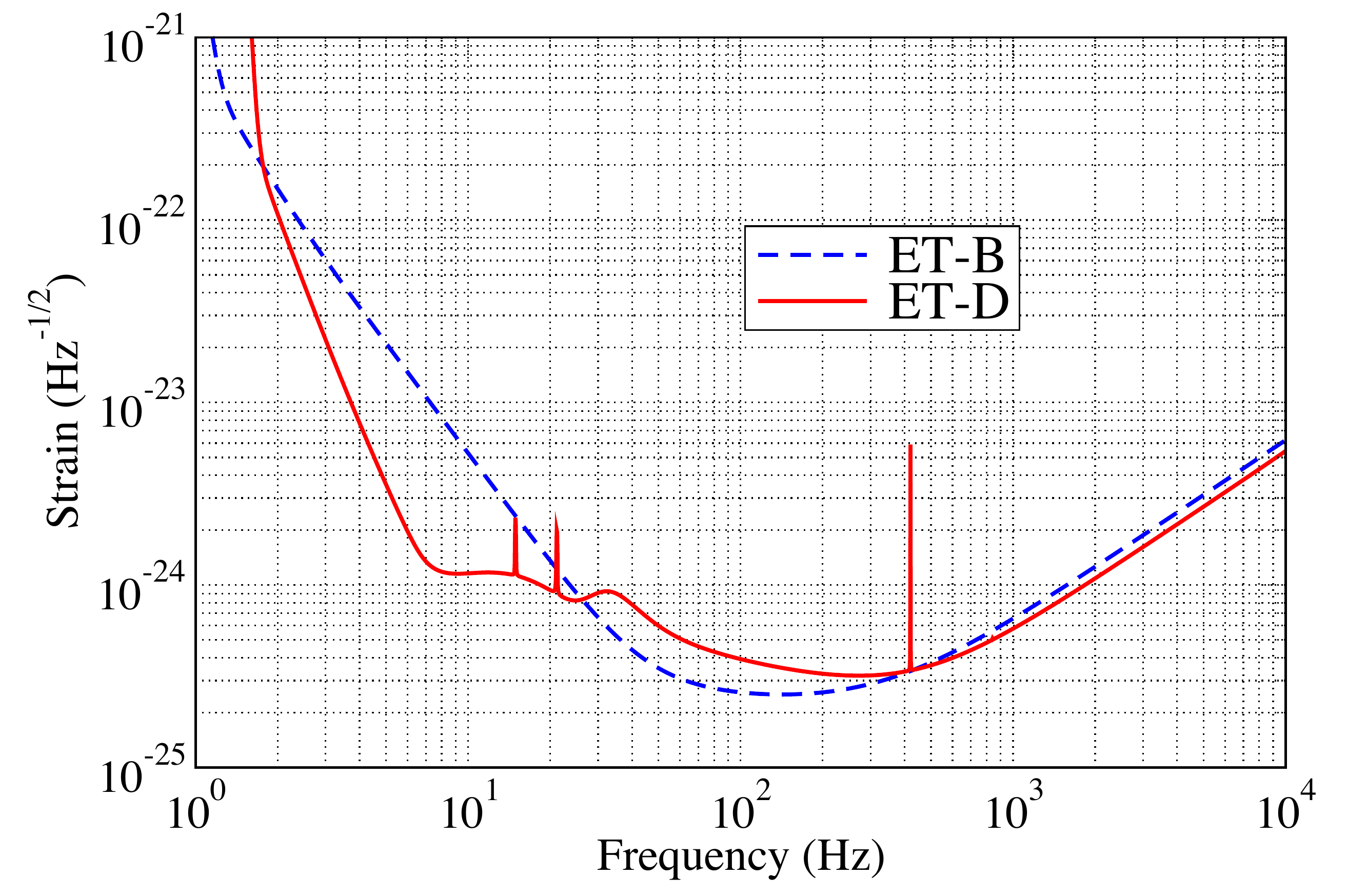}
\includegraphics[width=0.71\textwidth]{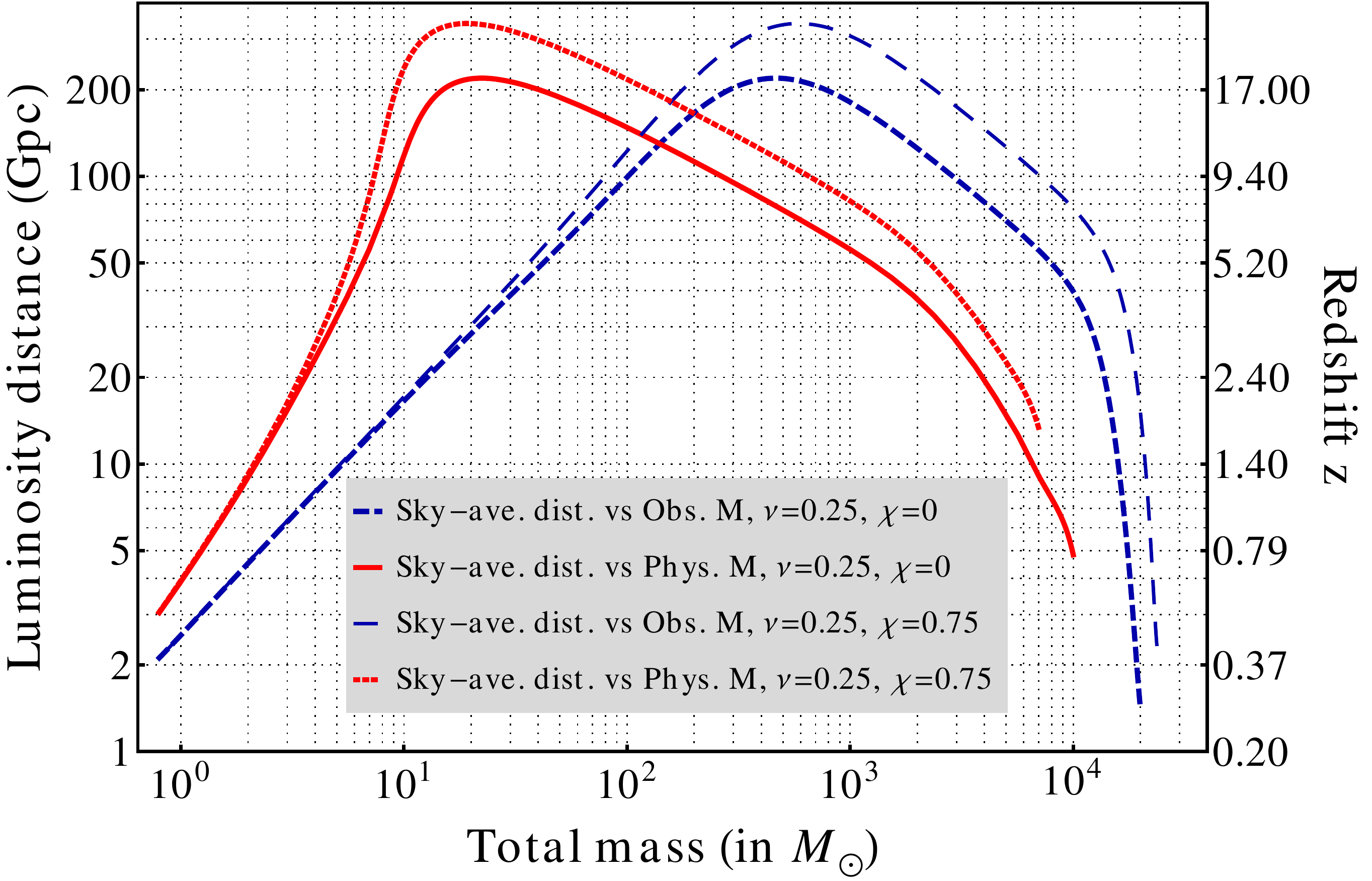}
\end{center}
\caption{\label{fig:sensitivity} The top panel shows ET's 
strain sensitivity for two optical configurations, ET-B \cite{ET-B} and ET-D 
\cite{Hild:2010id}.
The bottom panel plots ET-B's distance reach for compact binary
mergers as a function of the observed total mass (blue dashed curves) and
intrinsic total mass (red solid curves) for non-spinning binaries 
(lower curves) and binaries with dimensionless spins of $0.75$ (upper curves).}
\end{figure*}

Figure \ref{fig:sensitivity}, top panel, shows the sensitivity of 
each V-shaped detector in ET, for two different optical configurations.
The red solid curve, labelled ET-D, shows the sensitivity of a xylophone 
configuration \cite{Hild:2009ns} in which two interferometers are installed in each V of 
the triangle, one that has good high-frequency sensitivity and the other 
with good low-frequency sensitivity.  The blue dashed curve, labelled ET-B,
shows the best possible sensitivity for a single detector in each V 
of the triangle.  Compared to a single Michelson, the xylophone 
configuration improves the sensitivity by a factor of 2-10 in 
the frequency range 6-10 Hz. This improved low-frequency 
sensitivity makes ET-D a lot more attractive as it greatly
enhances the live time of stellar mass binaries in band and also makes it
possible to observe \ac{IMBH} binaries in the mass range 
100-$10^3\,M_\odot,$ at redshifts up to $z\sim 20,$ depending on 
their total mass, compared to ET-B's reach of $z\sim 5$-10.

The panel at the bottom of Figure \ref{fig:sensitivity} plots the distance 
reach of ET-B for compact binary signals as a function of the source's total 
mass \cite{DSD}.  More precisely, what is plotted is the distance at which a source with
random orientation, polarization and location on the sky would, on average, 
produce an SNR of 8. Blue dashed curves give the reach as a function of the 
{\em observed mass} of the source. The reach is shown for two archetypal systems,
long-dashed curve corresponds to systems with large spins, while short-dashed 
curve is the reach for non-spinning systems. Red solid curves plot the reach 
as a function of the {\em intrinsic mass} of the source.  They are obtained 
by shifting each point on the blue curve to the left by a factor $(1+z)$ to
account for the ``redshift'' of the system's total mass.

\subsection{ET mock data challenges}
Extrapolating the local rate of expected compact binary coalescences to
the distant Universe, a third generation detector like ET should detect 
millions of merging events per year 
\cite{Sathyaprakash:2009xt}.  At any one time, hundreds 
of overlapping signals could be present in the sensitive band of the detector. 
The number and variety of sources detected depends a lot on the nature of 
data analysis algorithms used in discriminating one signal from the 
other. ET poses real challenges in signal discrimination and
accurate measurement of the source parameters, the latter critical
if ET were to be used for precision cosmography and novel tests of \ac{GR}. 
The problem has begun to be addressed via ET mock data challenges. 

The first challenge, concluded recently \cite{Regimbau:2012ir}, tackled 
a limited set of questions of how effective are the pipelines currently used in
GW data analysis in discriminating overlapping signals in ET
and whether the population of sources causes a confusion background
degrading the detector sensitivity at lower frequencies.
We found that the population of overlapping sources in ET's 
sensitive band do not form a confusion background obscuring
foreground sources. However, the presence of the population,
its spectrum and gross properties of the underlying sources,
can be inferred by cross-correlating the three ET data
streams. The \texttt{iHOPE} pipeline \cite{Abbott:2009tt}, currently used by the
LIGO-Virgo collaboration to detect compact binary coalescences,
is already very effective in discriminating overlapping sources. 
Indeed, we found that \texttt{iHOPE}'s detection efficiency, at a
given redshift, was about $\sim 20\% $ smaller than that of an ideal coherent
search pipeline. Equivalently, the redshift reach of \texttt{iHOPE}
at 50\% efficiency is 10\% smaller than an ideal search pipeline.
This is expected as \texttt{iHOPE} is designed
for coincidence analysis and doesn't take full advantage of the
signal coherence in a detector network. Most interestingly, we found
that the null stream is a very powerful tool to identify stochastic
\ac{GW} backgrounds \cite{Regimbau:2012ir}. The residual 
formed by subtracting the power spectral densities of each detector
from that of the null stream clearly showed the presence of
the background compact binary population.

\subsection{Measuring the intrinsic masses of a binary}

It is well-known that GW interferometers can measure the total mass
and mass ratio of binaries to phenomenal accuracies \cite{AISS05}. 
For sources at cosmological distances the expansion of the Universe
causes the observed frequencies to be redshifted and so our detectors  
measure ``redshifted" masses, not the intrinsic masses. The 
observed mass is larger than the intrinsic mass by $(1+z),$ 
$M_{\rm obs}=M_{\rm int} (1+z).$ To infer the intrinsic mass it is necessary to 
know the source's cosmological redshift. It might not always be possible to
identify the host galaxy and directly measure its redshift, either because
the source is not well localized on the sky or because the host galaxy is too
far away.  Hence one is faced with the problem of having to infer the source's
redshift from the luminosity distance. 

Compact binary signals are standard sirens and our detectors can 
directly measure the source's luminosity distance $D_{\rm L}$, which, 
together with a cosmological model $D_{\rm L}(z;\, \Omega_\Lambda,\,
\Omega_{\rm M},\,w,\,\ldots),$ can, in principle, give the source's 
redshift. However, the luminosity distance is not 
measured very accurately due to its strong correlation with the source's
orientation and polarization. For sources with an SNR of 10, ET can
measure the distance to within 30\%. This means that the source's 
inferred redshift will be uncertain by a similar factor. Therefore,
the error in the determination of the intrinsic masses of a binary will 
be dominated by the uncertainty in the measurement of the luminosity distance.
Even so, ET should be able to measure the masses of most binaries to
within a factor of 2 --- an important factor in some of the scientific
objectives of ET\footnote{Note that ET's test of \ac{GR},
which requires accurate measurement of the system's masses and spins,
will not suffer from the redshift induced errors as they
test the orbital evolution of the source and are agnostic to whether
the masses are intrinsic or redshifted.}.

\section{ET's science objectives}

The goal of advanced detectors is to make the first detection of \acp{GW}
and establish the field of gravitational astronomy. Third generation detectors will be
sensitive to a greater variety of sources, sources at cosmological distances, signals with
large SNRs and so on.  Consequently, ET has a very impressive science potential 
and it will not be possible to cover every topic in any great detail.
We will, therefore, highlight an example each from cosmology, fundamental 
physics, nuclear physics and astrophysics and refer the reader to
the ET design study document for further details \cite{DSD}.

\subsection{Cosmology: Exploring black hole seeds}

The origin and evolution of \acp{BH} that seem to populate galactic
cores is one of the unsolved problems in modern cosmology. Seeds
of supermassive \acp{BH}  might have been initially very small (100's
to 1000's of $M_\odot$) and grew by accretion of gas and merger with 
other \acp{BH}  or perhaps they were already massive when they formed
and have undergone few mergers. Current 
observations are insufficient to pin down even the basic questions, when
did the first \acp{BH}  form, what was the spectrum of their masses,
how did they grow, and so on. ET should be able to provide answers to some of these
questions and constrain models of \ac{BH} formation and growth in
the early history of the Universe \cite{Gair:2010dx,AmaroSeoane:2009ui}.  
An \ac{IMBH} binary of intrinsic total mass of $500\,M_\odot$ at $z=2$
will appear in ET as a $1.5\times 10^3\,M_\odot$ binary, lasting
for about 14 s from 1 Hz until merger. It will have an SNR of 120 and
490 in ET-B and ET-D, respectively. The same system will appear twice 
as massive at $z=5,$ and produces an SNR of 28 and 190 in ET-B and ET-D.

ET has its best reach for stellar mass \acp{BBH}. Systems with
their total mass in the range 10-200~$M_\odot$ can be observed in
both ET-B and ET-D at redshift range of 9.5-17 (cf.\ Figure \ref{fig:sensitivity},
bottom panel). \ac{IMBH} binaries of mass 
$100$-$10^3\,M_\odot$ can be observed in the redshift range $z\sim 5$-10 in
ET-B and up to redshift of 20 in ET-D. 

Moreover, ET should be able to measure their total mass to an accuracy of
at least 50\%, even after accounting for the error introduced by the
conversion of the luminosity distance to redshift that is needed to
infer the intrinsic mass from the observed mass. Therefore, ET could
confirm or rule out hierarchical models \cite{Volonteri:2002vz}, according 
to which seed \acp{BH} are \acp{IMBH} which grow by accreting
gas and merging with other \acp{BH}. \ac{ET} will carry out a census of 
the \ac{BH} population in the mass range $[10,\,10^3]\,M_\odot$ 
throughout the Universe and study their evolution as a function of redshift.
If \acp{IMBH} form a significant population of
seeds, ET is arguably the best instrument to study them \cite{Gair:2010dx}.

\subsection{Fundamental physics: Testing gravity with black holes}

Nearly a hundred years after its formulation, \ac{GR}
continues to be the preferred theory of gravity. However, the theory is 
yet to be tested in strong gravitational fields that occur in the vicinity 
of \ac{BH} horizons. Gravitational wave observations of compact binaries 
could facilitate many such tests 
\cite{Arun:2008xf,AW09,Mishra:2010tp,Keppel:2010qu,Yunes:2011bk,Cornish:2011ys}. 
A new Bayesian approach \cite{Li:2011cg} to testing the post-Newtonian 
formula for the phasing of \acp{GW}, which is known to seven orders in 
perturbation theory \cite{BFIJ02,BDEI04,Bliving} beyond the quadrupole
formula, has shown that such tests should already be possible with advanced
detectors. A 10\% deviation in the ``tail effect" \cite{BS93,BSat94}, an effect that accounts
for scattering off \acp{GW} off the curved geometry in the
vicinity of the binary, from \ac{GR} would be easily discernible with a 
catalogue of just 15 \acp{BNS} observed with advanced detectors.
ET will be able to push this limit by several orders of magnitude with
the millions of systems that it could observe.

In addition to the inspiral phase, it should also be possible to use
the merger phase of \acp{BBH} to test strong field predictions 
of \ac{GR}.  The coalescence of a pair of \acp{BH} in a binary results in a single 
\ac{BH} that is initially highly deformed.  Deformed \acp{BH} 
emit gravitational radiation that consists of a superposition of, in 
principle, an infinitely large number of exponentially damped sinusoidal 
waves, called {\em quasi-normal modes} \cite{Berti:2009kk}. 
The no-hair theorem implies that the 
mode frequencies and time constants of an astronomical \ac{BH} 
should all be determined by just two parameters: \ac{BH}'s mass and its
spin magnitude.  Observation of  quasi-normal modes consistent with this
prediction would provide a smoking gun evidence of the presence of \acp{BH},
as no other body will have such a unique spectrum of modes \cite{BHspect04}.  Furthermore, 
by resolving two or more quasi-normal modes it might be possible to test 
strong field predictions of \ac{GR} \cite{BHspect04}. For instance, it is
possible, in principle, to measure the system's total mass before and
after merger and test if the mass lost to gravitational radiation is 
as predicted by \ac{GR}.

Until now such tests have largely remained speculative as no one knew the 
spectrum of modes that would be excited in a newly formed \ac{BH}. Recent 
work \cite{Berti:2007a,Kamaretsos:2011um,Kamaretsos:2011aa} used numerical 
simulations of non-spinning \ac{BH} binaries for an in-depth 
investigation of which modes are excited and what their amplitudes are.
The study showed that the amplitude of the different modes excited
in the process of merger depended on the mass ratio of the progenitor
binary and that it will be possible to infer the 
masses of the component stars that merged to form a \ac{BH} \cite{Kamaretsos:2011um}.
It will be interesting to see if a progenitor binary's component spins
can also be measured from a knowledge of the amplitude of various
modes.  To test \ac{GR} using quasi-normal modes, a Bayesian model 
selection approach has now been developed \cite{Gossan:2011ha}. A preliminary study
carried out using this approach shows that ET will be able to detect 6\% or more
departure from \ac{GR} of the frequency of the dominant quasi-normal mode 
excited during the merger of a pair of \acp{IMBH} at $z=1.$
Future work covering the full spectrum of the normal modes and using a 
population of detected events, instead of just one event, is necessary
to judge how good such tests are.

\subsection{Cosmography: Measuring the Universe with standard sirens}

One of the most spectacular aspects of compact binary signals is that
their amplitude is completely determined by \ac{GR}, without the
need for any complicated astrophysical modelling of their environments. 
Moreover, imprinted in the evolution of their phase is the absolute
luminosity of the source. Our detectors can, therefore, measure both the
apparent luminosity and absolute luminosity of a source as they are
related to the signal's amplitude and the rate at which the signal's
frequency changes, respectively. This means that compact binaries are
{\em standard sirens}\footnote{GW signals are referred to as {\em sirens,} 
as opposed to {\em candles,} due to their close analogy with audio.} and
it would be possible to measure the luminosity distance to their host
galaxies without any additional calibrators of distance \cite{Schutz86}.

The availability of a standard siren immediately raises the possibility
of their application in cosmography \cite{DaHHJ06,Nissanke:2009kt}. 
However, two hurdles have to be overcome
for a successful use of these sirens. Firstly, it is necessary to
measure the redshift of host galaxies and secondly it is imperative to
control the bias in the luminosity distance arising from weak gravitational
lensing of cosmological sources. Both of these have been addressed in
the context of ET and it has been shown that coincident observation of
shGRBs and \acp{GW} could be used to simultaneously measure 
the redshift and luminosity distance. With about 500 shGRBs, which ET
could observe over five years, it would be possible to infer the dark energy
equation of state parameter $w$ to within 
a percent or two \cite{Sathyaprakash:2009xt}. Moreover, the
tidal effects in \acp{NS} could facilitate a direct measurement
of the intrinsic masses and hence decipher the source's redshift 
\cite{Messenger:2011gi}. This is currently being explored and could 
greatly enhance ET's capability to measure cosmological parameters, as 
one can use the entire \ac{BNS} population
instead of a sub-set observed in coincidence with shGRBs.  

Del Pozzo has recently studied \cite{DelPozzo:2011yh} a statistical approach, proposed originally 
by Schutz \cite{Schutz86}, to measure the luminosity distance-redshift relation using a
population of \acp{BBH}. He has shown that such a population 
observed with advanced detectors could determine the Hubble 
constant to within a few percent.  This is a very encouraging result as
ET would observe millions of \acp{BBH} and such a large population
can be used to measure not just the Hubble constant but other cosmological 
parameters. A detailed study is needed to assess ET's ability to 
measure cosmological parameters with \acp{BBH}.

\subsection{Nuclear Physics: Probing neutron star cores}

Neutron star cores are laboratories of extreme conditions of density,
gravity and magnetic fields (for reviews on \acp{NS} and their
dynamics see \cite{Lattimer:2004pg,Chamel:2008ca,2011GReGr..43..409A}). 
The structure and composition of \ac{NS} cores have largely remained 
unresolved even half-century after pulsars were first discovered.
Their cores could be host to unknown physics and
might be composed of quark-gluon plasma, hyperons or other exotica 
\cite{Lattimer:2004pg,Chamel:2008ca}. 
Understanding the equation of state of \acp{NS} and the 
structure and composition of their cores, could provide deeper insights 
into fundamental nuclear physics, complimentary to heavy ion collision
experiments \cite{Ruan:2010vc}. 

Crust-core interaction in \acp{NS} involve vast amounts 
of energy that could generate transient bursts of GW. Such bursts 
would be observed in ET if the energy involved is $\gsim 10^{-12}\,M_\odot$ 
and the source is within the Milky Way, or if the energy is 
$\gsim 10^{-8}\,M_\odot$ and the source is not farther than the Andromeda
galaxy. Transient phenomena observed with radio, x-ray and gamma-ray telescopes
require that \acp{NS} are frequently converting such vast amounts
of energy into electromagnetic waves.
For example, sudden decrease in the rotation periods of \acp{NS}, 
so-called {\em glitches} \cite{Espinoza:2011pq}, in radio pulsars are believed to be 
caused by the exchange of energy between the differentially rotating
core and crust. Glitches in the Vela pulsar are quite frequent,
occurring once every few years. The largest of the glitches could involve 
$10^{-12}\,M_\odot.$ Similarly, giant explosions in highly magnetized
\acp{NS}, the so-called {\em magnetars} \cite{Arons2003} with magnetic fields of
$\sim 10^{15}$~G, require that isotropic emissions of $\sim 10^{46}\,
\rm erg$ could be involved.

Such star quakes might lead to normal mode oscillations of \ac{NS}
cores, resulting in characteristic oscillation modes. A wide range of
different modes are possible, their frequencies and damping times 
depending on the mass and radius of the star \cite{AnderssonKokkotas1998}. 
The nature of the modes depends on the restoring force in play and 
these could be fundamental- or f-modes, g-modes, p-modes, 
and so on \cite{AnderssonKokkotas1998}. By observing a particular mode,
it should be possible to solve for the star's mass and radius. In
turn, the relationship between the mass and radius of a star depends 
almost uniquely on the equation of state of the star, with little 
degeneracy amongst different equations of state. Thus, detection and 
identification of \ac{NS} normal modes could provide invaluable insight into \acp{NS}. 

More recently, several groups have pointed out that the inspiral phase
in the coalescence of \ac{BNS} systems could also be used to measure the
equation of state of \acp{NS} 
\cite{FlanaganHinderer2007,Read:2008pp,Read:2009bns,2011GReGr..43..409A,Lackey:2011vz}. 
It might actually be possible to
do this in two different ways. The finite size of \acp{NS} induces
a post-Newtonian correction to the phasing of \acp{GW}, called
tidal terms, which first occur at order $(v/c)^{10}$ beyond the quadrupole 
approximation \cite{FlanaganHinderer2007}. Currently, 
the first two tidal terms are known and should
be adequate for deciphering the equation of state. In addition to the
secular post-Newtonian tidal effect, the merger of \acp{NS} 
results in an unstable bar-like structure that spins at a frequency of
$1.5$ kHz and could last for tens of milliseconds \cite{2011GReGr..43..409A}. 
The precise nature of 
the bar-mode instability and the spectrum of emitted radiation depends
on the equation of state of \acp{NS} which ET could decipher from
events that occur within 100 Mpc.

\subsection{Astrophysics: Catching supernovae in their act}

First \ac{GW} detectors, resonant bar antennas, were
built solely to detect \acp{GW} from supernovae (SNe). They
are still sought after by current \ac{GW} detectors due to
the immense insight they could provide about the phenomenon \cite{bethe:90}.
It is expected that the gravitational collapse and the
ensuing explosion can be fully understood only 
by studying the deep interiors of the proto-neutron star that
forms in the process, which is 
inaccessible to electromagnetic observations. Modelling SNe
involves inputs from almost all branches of physics and current
simulations of the process are far from complete \cite{ott:09b}. 
Most of these models predict that the collapse could be quite non-axisymmetric and SNe 
could convert $\sim 10^{-7}$-$10^{-8}\,M_\odot$ into a burst of gravitational
radiation in the frequency range of 200-1000 Hz that lasts for 
10's of milliseconds \cite{ott:09b}.

Imprint in the gravitational radiation emitted during a SN is the 
dynamics of the SN engine of a dying massive star. The emitted
radiation could place strong constraints on the SN mechanism.
Advanced interferometers should be able to observe an event
occurring in the Milky Way or the Magellanic Clouds. However, even
the most optimistic estimates predict no more than a few events
per century. Thanks to a number of starburst galaxies, the rate 
increases to a few per decade within a distance of about 5 Mpc 
\cite{ando:05,strigari:05}, while ET's distance reach to SNe 
is about 2-4 Mpc.  Therefore, ET may see some SNe during its lifetime and would 
have the ability to provide strong hints for a particular SN mechanism or
evidence against another --- crucial astrophysics information 
that is unlikely to be attainable in other ways. 

SNe are conceivably the most interesting multi-messenger sources
that could be observed using optical, radio, x-ray and gamma-ray 
telescopes, neutrino detectors and \ac{GW} interferometers.
Transient astronomy, currently under vigorous development, will 
enable frequent all-sky surveys of transient phenomena and will 
regularly detect SNe in the nearby universe.  On a time-scale similar 
to ET, megaton-class water \v{C}erenkov neutrino detectors could be 
in operation with a distance reach of $\sim 5$~Mpc \cite{Kistler:2008us}.  Coincident 
observation of SNe in neutrino, optical and GW windows will provide 
astrophysical information that is critical to understanding a range 
of phenomena associated with SNe: stellar collapse, core-collapse
SNe, formation of \acp{BH} and \acp{NS} by gravitational
collapse, gamma-ray bursts, etc.  However, the astrophysics and 
physics information provided by \acp{GW} observed from 
a core collapse SNe with ET goes beyond this as the emitted radiation
carries information on the high-density nuclear equation of state, 
explosion asymmetries and \ac{NS} kicks, and can help uncover 
rare events such as the accretion-induced collapse of a white dwarf 
to a \ac{NS}, or silent SNe that have very weak electromagnetic signatures.

\ack
This work began with the support of the European 
Commission under the Framework Programme 7 (FP7) "Capacities", 
project \emph{Einstein Telescope} (ET) design study (Grant 
Agreement 211743), \texttt{http://www.et-gw.eu/.} We 
thank Drs Asad Ali and Evan Ochsner for comments.  

\section*{References}
\bibliography{ref-list}{}

\providecommand{\newblock}{}
\begin{thebibliography}{10}
\expandafter\ifx\csname url\endcsname\relax
  \def\url#1{{\tt #1}}\fi
\expandafter\ifx\csname urlprefix\endcsname\relax\def\urlprefix{URL }\fi
\providecommand{\eprint}[2][]{\url{#2}}

\bibitem{DSD}
Abernathy M {\em et~al.\/} 2011 {\em Einstein gravitational wave Telescope:
  Conceptual Design Study\/} (available from \em European Gravitational
  Observatory, \rm document number ET-0106A-10)

\bibitem{Gair:2010dx}
Gair J~R, Mandel I, Miller M and Volonteri M 2011 {\em Gen.Rel.Grav.\/} {\bf
  43} 485--518

\bibitem{AmaroSeoane:2009ui}
Amaro-Seoane P and Santamaria L 2010 {\em Astrophys.J.\/} {\bf 722} 1197--1206
  (\textit{Preprint} \eprint{0910.0254})

\bibitem{2009Natur.460...73F}
{Farrell} S~A, {Webb} N~A, {Barret} D, {Godet} O and {Rodrigues} J~M 2009 {\em
  Nature\/} {\bf 460} 73--75

\bibitem{Smith2009}
Smith (for~the LIGO Scientific~Collaboration) J 2009 {\em Classical and Quantum
  Gravity\/} {\bf 26} 114013

\bibitem{advLIGO:2007}
Abbott B {\em et~al.\/} 2007 Advanced ligo reference design Tech. Rep.
  {LIGO}-M060056-08-M {LIGO} Project
  \urlprefix\url{http://www.ligo.caltech.edu/docs/M/M060056-08/M060056-08.pdf}

\bibitem{0264-9381-23-19-S01}
Acernese F {\em et~al.\/} 2006 {\em Classical and Quantum Gravity\/} {\bf 23}
  S635--S642

\bibitem{LCGTProject}
Large-scale cryogenic gravitational-wave telescope project\\
  \urlprefix\url{http://www.icrr.u-tokyo.ac.jp/gr/LCGT.html}

\bibitem{Abadie:2010cf}
Abadie J {\em et~al.\/} (LIGO Scientific Collaboration, Virgo Collaboration)
  2010 {\em Class.Quant.Grav.\/} {\bf 27} 173001 (\textit{Preprint}
  \eprint{1003.2480})

\bibitem{Read:2009bns}
Read J~S {\em et~al.\/} 2009 {\em Phys. Rev.\/} {\bf D79} 124033
  (\textit{Preprint} \eprint{0901.3258})

\bibitem{2011GReGr..43..409A}
{Andersson} N, {Ferrari} V, {Jones} D~I, {Kokkotas} K~D, {Krishnan} B, {Read}
  J~S, {Rezzolla} L and {Zink} B 2011 {\em General Relativity and
  Gravitation\/} {\bf 43} 409--436

\bibitem{Nakar2007}
Nakar E 2007 {\em Physics Reports\/} {\bf 442} 166--236

\bibitem{Dietz:2010eh}
Dietz A 2011 {\em Astron.Astrophys.\/} {\bf 529} A97 (\textit{Preprint}
  \eprint{1011.2059})

\bibitem{Harry:2010fr}
Harry I~W and Fairhurst S 2011 {\em Phys.Rev.\/} {\bf D83} 084002
  (\textit{Preprint} \eprint{1012.4939})

\bibitem{Belczynski:2011qp}
Belczynski K, Bulik T, Dominik M and Prestwich A 2011  (\textit{Preprint}
  \eprint{1106.0397})

\bibitem{Belczynski:2010tb}
Belczynski K, Dominik M, Bulik T, O'Shaughnessy R, Fryer C {\em et~al.\/} 2010
  (\textit{Preprint} \eprint{1004.0386})

\bibitem{ShapiroEtAl09}
{Shapiro} C, {Bacon} D, {Hendry} M and {Hoyle} B 2009  (\textit{Preprint}
  \eprint{{arXiv:0907.3635}})

\bibitem{ChassandeMottin:2010zh}
Chassande-Mottin E, Hendry M, Sutton P~J and Marka S 2011 {\em Gen.Rel.Grav.\/}
  {\bf 43} 437--464

\bibitem{Freise:2008dk}
Freise A, Chelkowski S, Hild S, Del~Pozzo W, Perreca A {\em et~al.\/} 2009 {\em
  Class.Quant.Grav.\/} {\bf 26} 085012

\bibitem{Freise:2009nz}
Freise A, Hild S, Somiya K, Strain K, Vicere A {\em et~al.\/} 2011 {\em
  Gen.Rel.Grav.\/} {\bf 43} 537--567

\bibitem{Regimbau:2012ir}
Regimbau T, Dent T, Del~Pozzo W, Giampanis S, Li T~G {\em et~al.\/} 2012
  Submitted to Physical Review D -- 18 pages, 8 figures (\textit{Preprint}
  \eprint{1201.3563})

\bibitem{Wen:2005ui}
Wen L and Schutz B~F 2005 {\em Class.Quant.Grav.\/} {\bf 22} S1321--S1336
  (\textit{Preprint} \eprint{gr-qc/0508042})

\bibitem{Hild:2009ns}
Hild S, Chelkowski S, Freise A, Franc J, Morgado N {\em et~al.\/} 2010 {\em
  Class.Quant.Grav.\/} {\bf 27} 015003

\bibitem{ET-B}
Hild S, Chelkowski S and Freise A 2008  (\textit{Preprint}
  \eprint{arXiv:0810.0604})

\bibitem{Hild:2010id}
Hild S, Abernathy M, Acernese F, Amaro-Seoane P, Andersson N {\em et~al.\/}
  2011 {\em Class.Quant.Grav.\/} {\bf 28} 094013

\bibitem{Sathyaprakash:2009xt}
Sathyaprakash B, Schutz B and Van Den~Broeck C 2010 {\em Class.Quant.Grav.\/}
  {\bf 27} 215006 (\textit{Preprint} \eprint{0906.4151})

\bibitem{Abbott:2009tt}
Abbott B {\em et~al.\/} (LIGO Scientific Collaboration) 2009 {\em Phys.Rev.\/}
  {\bf D79} 122001 (\textit{Preprint} \eprint{0901.0302})

\bibitem{AISS05}
Arun K~G, Iyer B~R, Sathyaprakash B~S and Sundararajan P~A 2005 {\em
  Phys.~Rev.~D\/} {\bf 71} 084008 erratum-ibid. ~{\bf D } 72, 069903 (2005)
  (\textit{Preprint} \eprint{gr-qc/0411146})

\bibitem{Volonteri:2002vz}
Volonteri M, Haardt F and Madau P 2003 {\em Astrophys. J.\/} {\bf 582} 559--573
  (\textit{Preprint} \eprint{astro-ph/0207276})

\bibitem{Arun:2008xf}
Arun K, Mishra C, Broeck C~V~D, Iyer B, Sathyaprakash B {\em et~al.\/} 2009
  {\em Class.Quant.Grav.\/} {\bf 26} 094021 (\textit{Preprint}
  \eprint{0810.5727})

\bibitem{AW09}
Arun K~G and Will C~M 2009 {\em Class. Quant. Grav.\/} {\bf 26} 155002
  (\textit{Preprint} \eprint{arXiv: 0904.1190})

\bibitem{Mishra:2010tp}
Mishra C~K, Arun K, Iyer B~R and Sathyaprakash B 2010 {\em Phys.Rev.\/} {\bf
  D82} 064010

\bibitem{Keppel:2010qu}
Keppel D and Ajith P 2010 {\em Phys.Rev.\/} {\bf D82} 122001 (\textit{Preprint}
  \eprint{1004.0284})

\bibitem{Yunes:2011bk}
Yunes N 2011  (\textit{Preprint} \eprint{1112.3694})

\bibitem{Cornish:2011ys}
Cornish N, Sampson L, Yunes N and Pretorius F 2011 {\em Phys.Rev.\/} {\bf D84}
  062003 (\textit{Preprint} \eprint{1105.2088})

\bibitem{Li:2011cg}
Li T, Del~Pozzo W, Vitale S, Van Den~Broeck C, Agathos M {\em et~al.\/} 2011
  (\textit{Preprint} \eprint{1110.0530})

\bibitem{BFIJ02}
Blanchet L, Faye G, Iyer B~R and Joguet B 2002 {\em Phys. Rev. D\/} {\bf 65}
  061501(R) {Erratum-ibid~{\bf 71}, 129902(E) (2005)} (\textit{Preprint}
  \eprint{gr-qc/0105099})

\bibitem{BDEI04}
Blanchet L, Damour T, Esposito-Far{\`e}se G and Iyer B~R 2004 {\em Phys. Rev.
  Lett.\/} {\bf 93} 091101 (\textit{Preprint} \eprint{gr-qc/0406012})

\bibitem{Bliving}
Blanchet L 2006 {\em Living Rev. Rel.\/} {\bf 9} 4 (\textit{Preprint}
  \eprint{gr-qc/0202016})

\bibitem{BS93}
Blanchet L and Sch{\"a}fer G 1993 {\em Class. Quantum Grav.\/} {\bf 10}
  2699--2721

\bibitem{BSat94}
Blanchet L and Sathyaprakash B~S 1994 {\em Class. Quantum Grav.\/} {\bf 11}
  2807--2832

\bibitem{Berti:2009kk}
Berti E, Cardoso V and Starinets A~O 2009 {\em Class.\ Quant.\ Grav.\/} {\bf
  26} 163001 (\textit{Preprint} \eprint{0905.2975})

\bibitem{BHspect04}
Dreyer O, Kelly B, Krishnan B, Finn L~S, Garrison D and Lopez-Aleman R 2004
  {\em Class. Quantum Grav.\/} {\bf 21} 787 (\textit{Preprint}
  \eprint{gr-qc/0309007})

\bibitem{Berti:2007a}
Berti E, Cardoso J, Cardoso V and Cavagli\'a M 2007 {\em Phys. Rev.\/} {\bf
  D76} 104044 (\textit{Preprint} \eprint{0707.1202})

\bibitem{Kamaretsos:2011um}
Kamaretsos I, Hannam M, Husa S and Sathyaprakash B 2012 {\em Phys.Rev.\/} {\bf
  D85} 024018 (\textit{Preprint} \eprint{1107.0854})

\bibitem{Kamaretsos:2011aa}
Kamaretsos I 2011  See the article entitled: From black holes to their
  progenitors: A full population study in measuring black hole binary
  parameters from ringdown signals (\textit{Preprint} \eprint{1112.3077})

\bibitem{Gossan:2011ha}
Gossan S, Veitch J and Sathyaprakash B 2011  (\textit{Preprint}
  \eprint{1111.5819})

\bibitem{Schutz86}
Schutz B~F 1986 {\em Nature (London)\/} {\bf 323} 310

\bibitem{DaHHJ06}
Dalal N, Holz D~E, Hughes S~A and Jain B 2006 {\em Phys. Rev.\/} {\bf D74}
  063006 (\textit{Preprint} \eprint{astro-ph/0601275})

\bibitem{Nissanke:2009kt}
Nissanke S, Holz D~E, Hughes S~A, Dalal N and Sievers J~L 2010 {\em
  Astrophys.J.\/} {\bf 725} 496--514 (\textit{Preprint} \eprint{0904.1017})

\bibitem{Messenger:2011gi}
Messenger C and Read J 2011 {\em To appear in Phys.\, Rev.\, Lett\/}
  (\textit{Preprint} \eprint{1107.5725})

\bibitem{DelPozzo:2011yh}
Del~Pozzo W 2011  (\textit{Preprint} \eprint{1108.1317})

\bibitem{Lattimer:2004pg}
Lattimer J and Prakash M 2004 {\em Science\/} {\bf 304} 536--542
  (\textit{Preprint} \eprint{astro-ph/0405262})

\bibitem{Chamel:2008ca}
Chamel N and Haensel P 2008 {\em Living Rev.Rel.\/} {\bf 11} 10
  (\textit{Preprint} \eprint{0812.3955})

\bibitem{Ruan:2010vc}
Ruan L 2010 {\em Front.Phys.China\/} {\bf 5} 205--214 (\textit{Preprint}
  \eprint{1007.2882})

\bibitem{Espinoza:2011pq}
Espinoza C~M, Lyne A~G, Stappers B~W and Kramer M 2011  (\textit{Preprint}
  \eprint{1102.1743})

\bibitem{Arons2003}
Arons J 2003 {\em Astrophys.\ J.\/} {\bf 589} 871--892 (\textit{Preprint}
  \eprint{astro-ph/0208444})

\bibitem{AnderssonKokkotas1998}
Andersson N and Kokkotas K 1998 {\em Mon.\ Not.\ R.\ Astron.\ Soc.\/} {\bf 299}
  1059--1068 (\textit{Preprint} \eprint{gr-qc/9711088})

\bibitem{FlanaganHinderer2007}
{Flanagan} {\'E}~{\'E} and {Hinderer} T 2008 {\em Phys.\ Rev.\ D\/} {\bf 77}
  021502 (\textit{Preprint} \eprint{0709.1915})

\bibitem{Read:2008pp}
Read J~S, Lackey B~D, Owen B~J and Friedman J~L 2009 {\em Phys. Rev.\/} {\bf
  D79} 124032 (\textit{Preprint} \eprint{0812.2163})

\bibitem{Lackey:2011vz}
Lackey B~D, Kyutoku K, Shibata M, Brady P~R and Friedman J~L 2011
  (\textit{Preprint} \eprint{1109.3402})

\bibitem{bethe:90}
Bethe H~A 1990 {\em Rev. Mod. Phys.\/} {\bf 62} 801--866

\bibitem{ott:09b}
Ott C~D 2009 {\em Class.Quant.Grav.\/} {\bf 26} 204015 (\textit{Preprint}
  \eprint{0905.2797})

\bibitem{ando:05}
{Ando} S, Beacom F and Y\"uksel H 2005 {\em Phys. Rev. Lett.\/} {\bf 95} 171101

\bibitem{strigari:05}
{Strigari} L~E, {Beacom} J~F, {Walker} T~P and {Zhang} P 2005 {\em JCAP\/} {\bf
  4} 17

\bibitem{Kistler:2008us}
Kistler M~D, Yuksel H, Ando S, Beacom J~F and Suzuki Y 2011 {\em Phys.Rev.\/}
  {\bf D83} 123008 (\textit{Preprint} \eprint{0810.1959})

\end{thebibliography}

\end{document}